\documentclass[12pt,a4paper,hypertex]{article}
\usepackage{jheppub}

\allowdisplaybreaks[1]
\usepackage{amsmath,bm}
\usepackage{subfigure}
\usepackage{color}
 \usepackage{ulem}

\begin{document}

\title{
Self-Consistent  Large-$N$ Analytical Solutions of Inhomogeneous Condensates 
in Quantum ${\mathbb C}P^{N-1}$ Model
} 
\author[1,2]{Muneto Nitta} 
\author[2]{and Ryosuke Yoshii} 
\affiliation[1]{Department of Physics,  Keio University, 4-1-1 Hiyoshi, Kanagawa 223-8521, Japan}
\affiliation[2]{Research and Education Center for Natural Sciences, Keio University, 4-1-1 Hiyoshi, Kanagawa 223-8521, Japan}
\emailAdd{nitta@phys-h.keio.ac.jp}
\emailAdd{ryoshii@phys.keio.ac.jp}

\date{\today}
\abstract{
We give, for the first time, 
self-consistent large-$N$ analytical 
solutions of inhomogeneous condensates 
in the quantum ${\mathbb C}P^{N-1}$ model 
in the large-$N$ limit. 
We find a map from a set of gap equations of the ${\mathbb C}P^{N-1}$ model 
to those of the Gross-Neveu (GN) model 
(or the gap equation and the Bogoliubov-de Gennes equation), 
which enables us to find the self-consistent solutions. 
We find that the Higgs field of the ${\mathbb C}P^{N-1}$ model 
is given as a zero mode of solutions of the GN model, 
and consequently only topologically nontrivial solutions 
of the GN model yield nontrivial solutions 
of the ${\mathbb C}P^{N-1}$ model. 
A stable single soliton is constructed from 
an anti-kink of the GN model and 
has a broken (Higgs) phase inside its core,
in which ${\mathbb C}P^{N-1}$ modes are localized,
with a symmetric (confining) phase outside.
We further find a stable periodic soliton lattice 
constructed from a real kink crystal in the GN model,
while the Ablowitz-Kaup-Newell-Segur hierarchy 
yields multiple solitons at arbitrary separations. }

\maketitle

\section{Introduction}
Nonlinear sigma models 
such as the ${\mathbb C}P^{N-1}$ model in 1+1 dimensions
\cite{Eichenherr:1978qa, Golo:1978dd, Cremmer:1978bh} 
are known to share a number of phenomena common with 
3+1 dimensional QCD, 
e.g.\, asymptotic freedom, dynamical mass generation, confinement, and instantons 
\cite{Polyakov:1975rr, Polyakov:1975yp, Bardeen:1976zh, Brezin:1976qa, DAdda:1978vbw, DAdda:1978etr, Witten:1978bc,Novikov:1984ac}. 
The mass gap can be best shown in the large-$N$ analysis 
in which one solves the gap equations self-consistently, 
to be consistent with 
the Coleman-Mermin-Wagner (CMW) theorem 
forbidding a gapless excitations in 1+1 dimensions \cite{Coleman:1973ci,Mermin:1966fe}. 
The ${\mathbb C}P^{N-1}$ model, 
or the ${\mathbb C}P^1$ model 
equivalent to the $O(3)$ sigma model, appears in 
a wide range of physics from particle physics to 
condensed matter physics. 
The relation between the 1+1 dimensional Heisenberg antiferromagnetic spin chain and the $O(3)$ sigma model has been shown in Ref.\ \cite{Haldane:1982rj, Affleck}. 
Recently, the quantum phase transition, so-called deconfined criticality is proposed in the antiferromagnetic system \cite{Senthil, Nogueira}. 
The sigma model with topological term is known to describe the integer quantum Hall effect \cite{Pruisken:1984ni}. 
The supersymmetric ${\mathbb C}P^{N-1}$ model was also investigated 
\cite{Witten:1977xn, DiVecchia:1977nxl} 
for which the all order calculation in coupling constant is possible for Gell-Mann-Low function 
\cite{Novikov:1984ac}, 
and 
dynamical mass gap was proved by the mirror symmetry \cite{Hori:2000kt}. 
The analogy between 3+1 dimensional Yang-Mills theory and 1+1 dimensional sigma model, 
pointed out in Ref.\ \cite{Polyakov:1975rr}, 
has been recently revealed in a rather nontrivial way; 
a non-Abelian vortex string in a $U(N)$ gauge theory 
with $N$ scalar fields in the fundamental representation 
carries ${\mathbb C}P^{N-1}$ moduli \cite{Hanany:2003hp, Auzzi:2003fs, Eto:2005yh}  
(see Refs.~\cite{Tong:2005un,Eto:2006pg,Shifman:2007ce,Tong:2008qd}
 as a review),  
yielding a nontrivial relation between 
the ${\mathbb C}P^{N-1}$ model on 
the string worldsheet 
and the bulk gauge theory 
\cite{Hanany:2004ea,Shifman:2004dr}.
The ${\mathbb C}P^{N-1}$ model defined 
on an interval 
\cite{Milekhin:2012ca,Bolognesi:2016zjp}
or on a ring 
\cite{Monin:2015xwa} was also studied. 
The ${\mathbb C}P^{N-1}$ model or the $O(3)$ sigma model at finite temperature and/or density was also investigated in which Berezinskii- Kosterlitz-Thouless transition at nonzero density was examined \cite{Bruckmann:2016txt}. 
One of recent developments is
a resurgent structure of 
the ${\mathbb C}P^{N-1}$ model \cite{Dunne:2012ae,Misumi:2014jua}, 
in which a molecule of fractional instantons \cite{Eto:2004rz,Bruckmann:2007zh}
called a bion, plays a crucial role. 
In spite of tremendous studies of the ${\mathbb C}P^{N-1}$ model, 
there was no study on inhomogeneous configurations (such as solitons) 
 at quantum level,except for a numerical study of  
the ${\mathbb C}P^{N-1}$ model on an interval \cite{Bolognesi:2016zjp}.

The situation is rather different for an interacting fermionic theory:
the Gross-Neveu (GN) 
\cite{Gross:1974jv}
or Nambu-Jona-Lasino model 
\cite{Nambu:1961tp}, 
exhibiting dynamical symmetry breaking of 
discrete or continuous chiral symmetry,
thereby sharing an important property with QCD 
\cite{Dashen:1975xh, Shei:1976mn, Feinberg:1996gz}.  
This model is equivalent at the large-$N$ limit or 
in the mean field approximation 
to a set of 
the Bogoliubov-de Gennes (BdG) equations 
and the gap equation, 
appearing in condensed matter systems 
such as conducting polymers 
\cite{Takayama:1980zz, Brazovskii, Chodos}, 
superconductors, superfluids 
and ultracold atomic gases \cite{Machida:1984zz, Giorgini:2008zz,Yoshii:2011yt}.
Self-consistent analytical solutions such as a real kink \cite{Dashen:1975xh,Takayama:1980zz},
a twisted (complex) kink \cite{Shei:1976mn}, 
a real kink-anti-kink (polaron) \cite{Dashen:1975xh,Campbell:1981dc}, 
a real kink-anti-kink-kink \cite{Okuno:1983,Feinberg:1996gz,Feinberg:2002nq}
and more general real solutions \cite{Feinberg:2003qz}
have been known. 
Recently, a theoretical progress has been achieved for inhomogeneous condensates in the 1+1 dimensional (chiral) GN model, e.g.,\ the exact self-consistent and inhomogeneous condensates such as 
a real kink crystal \cite{Thies:2003br} 
(Larkin-Ovchinnikov(LO) state \cite{larkin:1964zz}), 
a chiral spiral (Fulde-Ferrell(FF) state \cite{Fulde:1964zz}), 
and a twisted kink crystal \cite{Basar:2008im} 
(FF-LO state) 
 have been found by mapping the equations to the nonlinear Schr\"odinger equation, 
and such states have been shown to be 
ground states in a certain region of 
the phase diagram for finite temperature and density 
\cite{Basar:2009fg}. 
More generally, multiple twisted kinks 
with arbitrary phase and positions 
 \cite{Takahashi:2012pk} 
can be further 
constructed systematically 
due to the integrable structure behind the model 
known as the Ablowitz-Kaup-Newell-Segur (AKNS) hierarchy for the nonlinear Schr\"odinger equation 
\cite{Ablowitz:1974ry,Correa:2009xa, Takahashi:2012aw}. 
Recent developments include  
time-dependent soliton scatterings \cite{Dunne:2013xta}, 
multi-component condensates 
\cite{Takahashi:2015nda,Thies:2015dim}, 
a ring geometry \cite{Yoshii:2014fwa},
and an interval with a Casimir force \cite{Flachi:2017cdo}.

In the present work, 
we reveal an unexpected relation between 
these two completely different theories, 
the ${\mathbb C}P^{N-1}$ and GN models developed independently. 
By finding a map from a set of gap equations of the ${\mathbb C}P^{N-1}$ model 
to those of the GN model,
we find self-consistent analytical solutions of stable inhomogeneous condensates 
in the quantum ${\mathbb C}P^{N-1}$ model, that is, 
a single soliton,  a soliton lattice 
and multiple solitons at arbitrary separations.

\section {Model and method}
We consider the  ${\mathbb C}P^{N-1}$ model on an infinite space: 
\begin{equation}
S=\int dtdx \left[(D_\mu n_i)^\ast (D^\mu n_i)-\lambda (n_i^\ast n_i-r)\right], 
\label{action}
\end{equation}
where $n^i$ ($i=1,\cdots, N$) are complex scalar fields, $D_\mu=\partial_\mu-iA_\mu$, and $\lambda(x)$ is a Lagrange multiplier. 
The ``radius" $r$ is known to have connection with a coupling constant $g_{\rm YM}$ in the Yang-Mills theory; $r=4\pi/g_{\rm YM}^2$ 
if we realize this model on a non-Abelian vortex 
in $U(N)$ gauge theory.
Here we note that the model does not have kinetic term for $A_\mu$ and thus we focus on the case of $A_\mu=0$ throughout this paper. 
We separate $n_i$ fields into a classical field $n_1=\sigma$ (real) and $n_i=\tau_i$ ($2, \cdots, N$).   
Integrating out the $\tau_i$ fields, we obtain the effective action for $\sigma$ as 
\begin{align}
S_{\rm eff}=\int dt dx \left[(N-1) \mathrm{Tr} \ln (-\partial_\mu \partial^\mu +\lambda)
+\partial_\mu\sigma\partial^\mu\sigma-\lambda (\sigma^2-r)\right]. 
\label{effaction}
\end{align}
In the following we consider and the leading contribution of $1/N$ expansion 
and thus we replace $N-1$ to $N$ 
\footnote{We note that the large $N$ limit is considered to obtain the self-consistent equations 
and the rest does not rely on the large $N$. 
Furthermore, the mean field approximation (for finite $N$) also yields the same self-consistent equations. 
Thus the results in the following are expected to be qualitatively correct even in the case of finite $N$. 
}. 
One can formally write down the total energy functional as 
\begin{equation}
E=N\sum_n\omega_n+\int dx \left[(\partial_x\sigma)^2+\lambda(\sigma^2-r)\right]. \label{eq:energy0}
\end{equation}
The corresponding gap equations obtained from the static condition with respect to $\lambda$ and $\sigma$ are \cite{Bolognesi:2016zjp} 
\begin{align}
\frac{N}{2}\sum_n \frac{f_n^2}{\omega_n}+\sigma^2-r=0, \label{gapeq}\\
\partial_x^2\sigma-\lambda\sigma=0, \label{eqsigma}
\end{align}
respectively, 
where $f_n(x)$ and $\omega_n$ are orthonormal eigenstates and eigenvalues of the following equation 
\begin{equation}
(-\partial_x^2 +\lambda)f_n(x)=\omega_n^2 f_n(x). \label{eigeneq}
\end{equation}
We need to solve Eqs.~(\ref{gapeq})--(\ref{eigeneq}) in a self-consistent manner. 
We here note from Eqs.~(\ref{eqsigma}) and (\ref{eigeneq}) that 
$\sigma$ is proportional to a zero mode $f_0$. 

It is well known that 
assuming a uniform state in infinite system, 
one finds the confining (unbroken) phase 
with a constant $\lambda$ to be a unique solution, 
to be consistent with the CMW theorem. 
For the case of a ring, in addition to it, there is 
a Higgs (broken) phase with a constant $\sigma$ for a smaller ring 
\cite{Monin:2015xwa}.

One of the main results of this paper is a map from those equations to the gap equation and eigenvalue equation for the GN model. 
In order to reduce the number of equations, we introduce the new field $\Delta$ such as 
\begin{equation}
\Delta^2+\partial_x\Delta=\lambda(x). 
\label{deflambda}
\end{equation}
By using this function, we find a solution to Eq.\ (\ref{eqsigma}): 
\begin{align}
\sigma=A\exp \left[\int^x dy \Delta(y)\right], \label{sigmasol}
\end{align}
where $A$ is the integral constant. 
The energy in Eq.~(\ref{eq:energy0}) can be rewritten as 
\begin{equation}
E_{\mathrm{tot}}=N\sum_n\omega_n-r\int^\infty_{-\infty} dx (\Delta^2+\partial_x\Delta)
+ \left.\sigma\partial_x \sigma\right|_{-\infty}^{\infty}.\label{eq:energy}
\end{equation}

The rather nontrivial step is to rewrite Eq.\ (\ref{eigeneq}) as [See Appendix A]
\begin{align}
\left(
\begin{array}{cc}
0 & \partial_x +\Delta \\
-\partial_x+\Delta & 0
\end{array}
\right)
\left(
\begin{array}{c}
f_n \\
g_n
\end{array}
\right)
=
\omega_n
\left(
\begin{array}{c}
f_n \\
g_n
\end{array}
\right),
\label{eqBdG}
\end{align}
where $g_n$'s are auxiliary fields and the elimination of $g_n$ yields Eq.\ (\ref{eigeneq}). 
We note that Eq.~(\ref{eqBdG}) together with Eq.\ (\ref{deflambda}) describes a supersymmetric quantum mechanics, 
in which the potential $\lambda$ is given by the superpotential $\Delta$ \cite{Cooper:1994eh}. 
Eq.\ (\ref{eqBdG}) is the positive energy part of the BdG or Andreev equation which corresponds to the Hartree-Fock equation of the GN model with $N$ flavors [See Appendix B]
\begin{equation}
L_{\mathrm{GN}}=\bar \psi i / \hspace{-0.5em}\partial \psi+\frac{g^2}{2}(\bar\psi \psi)^2. 
\end{equation}
The corresponding Hartree-Fock equation becomes $H\psi=E\psi$, with $H=-i\gamma^5 \partial_x+\gamma^0 \Delta$, where $\gamma^5=-\sigma_2$ and $\gamma^0=\sigma_1$ with the Pauli matrices $\sigma_i$. 
Here $\Delta$ (real) satisfies $\langle \bar\psi \psi\rangle=-\Delta/g$, which is called a gap equation.
It is known that the ${\mathbb Z}_2$ symmetry 
is spontaneously broken in the GN model,
yielding two discrete vacua. 

With a help of $g_n=(-\partial_x+\Delta)f_n/\omega_n$, one can show that $g_n$ automatically gives a orthonormal set if $f_n$ gives a orthonormal set. 
Eq.~(\ref{eqBdG}) has the particle-hole symmetry which enables us to obtain the set $\{-\omega_n, \tilde f_n, \tilde g_n\}$ from the set $\{\omega_n, f_n, g_n\}$ by $\tilde f_n=f_n$ and $\tilde g_n=-g_n$. 
By taking the derivative of Eq.\ (\ref{gapeq}) with respect to $x$ and by substituting Eqs.\ (\ref{sigmasol}) and $\omega_n g_n =(-\partial_x+\Delta)f_n$ into that, we obtain 
\begin{equation}
\Delta=\frac{N}{2r}\sum_n f_n g_n=-\frac{N}{2r}\sum_n \tilde f_n \tilde g_n,
\label{gapeqGN}
\end{equation}
which has the same form with the gap equation for the GN model. 
Here we note that corresponding fermionic coupling $Ng^2= N/2r$
 is proportional to the 't Hooft coupling in an underlying $U(N)$ gauge theory $Ng^2_{\mathrm{YM}}$.
Since we solve the differentiated one 
instead of Eq.\ (\ref{gapeq}) itself, we need to fix the integration constant $A$ for $\sigma$ by substituting Eq.\ (\ref{sigmasol}) into Eq.\ (\ref{gapeq}). 
For the BdG equation (\ref{eqBdG}) and gap equation (\ref{gapeqGN}), various exact self-consistent solutions are already known. 
From Eq.\ (\ref{eqBdG}) one can immediately find the zero mode solution 
\begin{equation} 
f_0(x)\propto \exp \left[\int^x dy \Delta(y)\right], 
\end{equation}
where the corresponding auxiliary field is $g_0(x)=0$. 
The zero mode solutions $f_0$ in 
the ${\mathbb C}P^{N-1}$ and GN models are identical.
As denoted below Eq.~(\ref{eigeneq}), 
the Higgs field $\sigma(x)$ in 
the ${\mathbb C}P^{N-1}$ model 
is proportional to the zero mode, 
thereby exists only when corresponding 
$\Delta$ in the GN model is topologically nontrivial with allowing 
a normalizable zero mode 
\cite{Jackiw:1975fn}

\section{Self-consistent analytical solutions}
In the GN model, a constant gap $\Delta=m$ is a solution  
which can be called the Bardeen-Cooper-Schrieffer (BCS) phase, 
whereas that for $m=0$ is called a normal phase. 
We show that the BCS and normal phases in the GN model 
correspond to the confining and 
Higgs phases in the ${\mathbb C}P^{N-1}$ model, respectively. 
For the constant solution, $\omega_n=\sqrt{(\pi n/L)^2+m^2}$ and the degenerated eigenfunctions are  $f_n^{(1)}=\sqrt{2} \sin \pi n x/L, f_n^{(2)}=\sqrt{2} \cos \pi n x/L$. 
For both the cases, $g^{(i)}_n(x)=(-\partial_x +m)f_n/\omega_n$ ($i=1,2$). 
Here we consider the periodic boundary condition in domain $[-L/2, L/2]$. 
The infinite system can be obtained by taking the proper limit of $L\rightarrow\infty$. 
The substitution $\Delta=m$ and corresponding eigenstates into Eq.\ (\ref{gapeqGN}) yields 
\begin{equation}
m=\frac{N}{r}\sum_n\frac{m}{\omega_n}, 
\label{ren1}
\end{equation}
while Eq.\ (\ref{gapeq}) becomes 
\begin{equation}
\sigma^2=r-N\sum_n\frac{1}{\omega_n}.  
\label{ren2}
\end{equation}
We find that the condition (\ref{ren1}) for $m\neq 0$ and (\ref{ren2}) for $\sigma=0$ are equivalent 
\begin{equation}
1=\frac{N}{r}\sum_n\frac{1}{\omega_n},
\label{rencoupling} 
\end{equation}
which gives the well known renormalization condition of the coupling constant $g^2=4\pi/r$. 
This results in two possibilities 
$\{\lambda=m^2, \sigma=0\}$ (confining phase) and 
$\{\lambda=0, \sigma=\mathrm {const}\}$ (Higgs phase), 
but only the former satisfies the gap equation (\ref{gapeq}) 
and the latter is not allowed in the infinite system  \cite{Milekhin:2012ca, Monin:2015xwa}.

\begin{figure}
\begin{center}
\includegraphics[width=20pc]{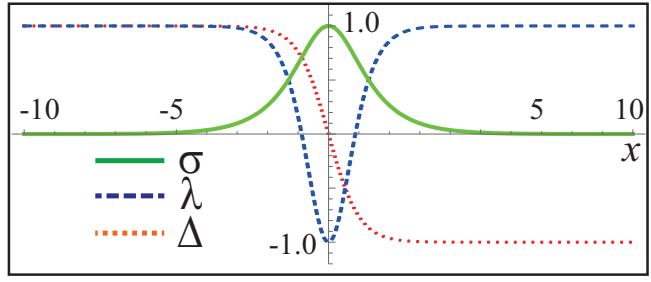}
\caption{The configuration of $\sigma$ (solid line) and $\lambda$ (dashed line) for $\Delta=-m\tanh mx$ (dotted line). 
Here we normalize as $\sigma(0)=1$ and $m=1$. }
\label{FigTanh}
\end{center}
\end{figure}

The solution $\Delta=-m\tanh mx$ is known as a topological kink solution 
interpolating two discrete vacua of the GN model,  
which has a zero mode localized near the kink. 
In the case of kink solution, the eigenvalue is the same with the constant solution $\omega_n=\sqrt{(\pi n/L)^2+m^2}$ while the degenerated eigenfunctions are $f_n^{(i)}=(\partial_x-m\tanh mx)g_n^{(i)}/\omega_n$ with $g_n^{(1)}=\sqrt{2} \sin \pi n x/L$, $g_n^{(2)}=\sqrt{2} \cos \pi n x/L$. 
We also have a normalizable zero mode $f_0(x)\propto 1/\cosh mx$, $g_0(x)=0$. 
Thus Eq.\ (\ref{gapeqGN}) yields
\begin{equation}
-m\tanh mx=\frac{N}{r}\sum_n\frac{-m\tanh mx}{\omega_n}, 
\label{gapeqtanh}
\end{equation}
which indeed gives the same condition with Eq.\ (\ref{ren1}). 
On the other hand, Eq.\ (\ref{gapeq}) implies
\begin{equation}
\sigma^2=r-N\sum_n\frac{1}{\omega_n}+\frac{m^2}{\cosh^2 mx} N \sum_n\frac{1}{\omega_n^3}.  
\end{equation}
In the case of $m\neq 0$, Eq.\ (\ref{gapeqtanh}) yields Eq.\ (\ref{rencoupling}) and we reach at 
\begin{equation}
\sigma=\frac{m}{\cosh mx}\sqrt{ N\sum_n\frac{1}{\omega_n^3}},  
\label{sigmasoltanh}
\end{equation}
which has a bright solitonic profile. 
Again, it is indeed proportional to the zero mode solution. 
In this case, the mass gap function becomes 
\begin{equation}
\lambda(x)=m^2 (1-2\cosh^{-2}mx), \label{eq:single-lambda}
\end{equation}
which has a gray soliton configuration and is called the P\"oschl-Teller potential \cite{Cooper:1994eh}. 
Since all the eigenenergies of this solution are non-negative, 
the solution is stable. 
In Fig.\ \ref{FigTanh}, we plot the configuration of $\sigma(x)$ and the mass gap function $\lambda(x)$. 
The energy of the soliton can be calculated 
by the energy $E_s$ for the soliton configuration in Eqs.~(\ref{sigmasoltanh}) and (\ref{eq:single-lambda})
subtracted by $E_0$ for the confining phase ($\sigma_0=0$ and $\lambda_0=m^2$), 
for both of which the third term in Eq.~(\ref{eq:energy}) vanishes from the equation of motion (\ref{eqsigma}) and $\omega_n$'s are the same. 
We thus obtain  
\begin{equation}
E_s-E_0=\int^\infty_{-\infty} dx r(\lambda_0 - \lambda_s)=4 r m.
\end{equation}

\begin{figure}
\begin{center}
\includegraphics[width=30pc]{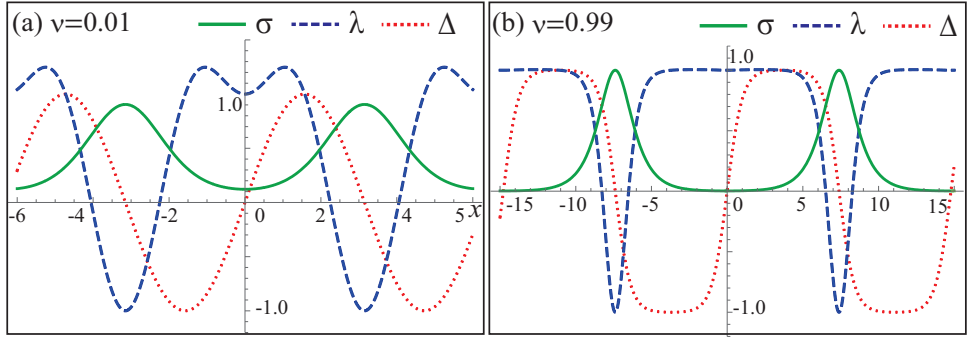}
\caption{The bright soliton lattice configuration of $\sigma$ (solid line) and $\lambda$ (dashed line) for  $\nu=10^{-2}$ (left figure) and $\nu=1-10^{-2}$ (right figure). 
The auxiliary field $\Delta$ (dotted line) are also plotted. 
Here we set, $m=1$ and normalize the peak of $\sigma$ to be $1$. }
\label{FigSN}
\end{center}
\end{figure} 

Since $\sigma$ has a localized profile function,
soliton core is in the Higgs (broken) phase 
where  
the ${\mathbb C}P^{N-1}$ modes are localized, 
while the bulk is in the confining (symmetric) phase, 
in contrast to a uniform system allowing only 
the confining phase in infinite system to be consistent with 
the CMW theorem. 
It is known that the correlation function behaves at large distance as $x^{-1/N}$ in 1+1 dimension \cite{Witten:1978qu}, 
which inhibits the long-range order for finite $N$. 
Here we have obtained the Higgs phase localized with length $\sim 1/m$, 
thus the robustness of our solitonic solution is expected if $N$ is sufficiently large as $\ln(1/m)\ll N$ \footnote{ It is also the case of the ${\mathbb C}P^{N-1}$ model on a ring:  
The Higgs phase is allowed for a smaller ring \cite{Monin:2015xwa}.
}.

The above solutions can be obtained from 
a soliton lattice obtained from a real kink crystal 
in the GN model: 
\begin{equation}
\Delta(x)= m \mathrm{sn}(mx,\nu), 
\label{snsol}
\end{equation}
where $\text {sn}$, $\text {cn}$, and $\text{dn}$ (appearing later) are the Jacobi functions and $\nu$ is elliptic parameter. 
Here the periodicity of the above solution is given by $\ell=4K(\nu)/m$, where $K(\nu)$ is a complete elliptic integral of the first kind. 
This solution together with Eq.\ (\ref{sigmasol}) gives a soliton lattice:
\begin{equation}
\sigma=A\left[\frac{-\sqrt{\nu}\mathrm{cn}(x,\nu)+\mathrm{dn}(x,\nu)}{1-\sqrt{\nu}}\right]^{\pm \frac{1}{\sqrt{\nu}}}. 
\end{equation}
In Fig.\ \ref{FigSN}, we plot the mass gap function $\lambda$ and $\sigma$ for  $\nu=10^{-2}$ and $\nu=1-10^{-2}$. The auxiliary field $\Delta$ are also plotted. 
The Higgs field $\sigma$ in this solution has a bright soliton lattice profile. 
By taking $\nu=1$ limit for $\Delta=m\mathrm{sn}(mx+K(\nu), \nu)$, $\lambda$ becomes constant and $\sigma=0$ in the whole system. 
This limit corresponds to the constant solution discussed above. 
On the other hand, $\Delta=m\mathrm{sn}(mx+2K(\nu), \nu)$ reduces to $\lambda=m[1-2/\cosh^2mx]$ and $\sigma(x)\propto m/\cosh mx$. 
This corresponds to the kink solution [See Appendix A]. 
Our periodic soliton solutions can be put on a ring,
while the previous studies on the ${\mathbb C}P^{N-1}$ model on a ring 
dealt with only constant configurations \cite{Monin:2015xwa}.

\section{Higher order self-consistent analytical solutions} 
\begin{figure}
\begin{center}
\includegraphics[width=30pc]{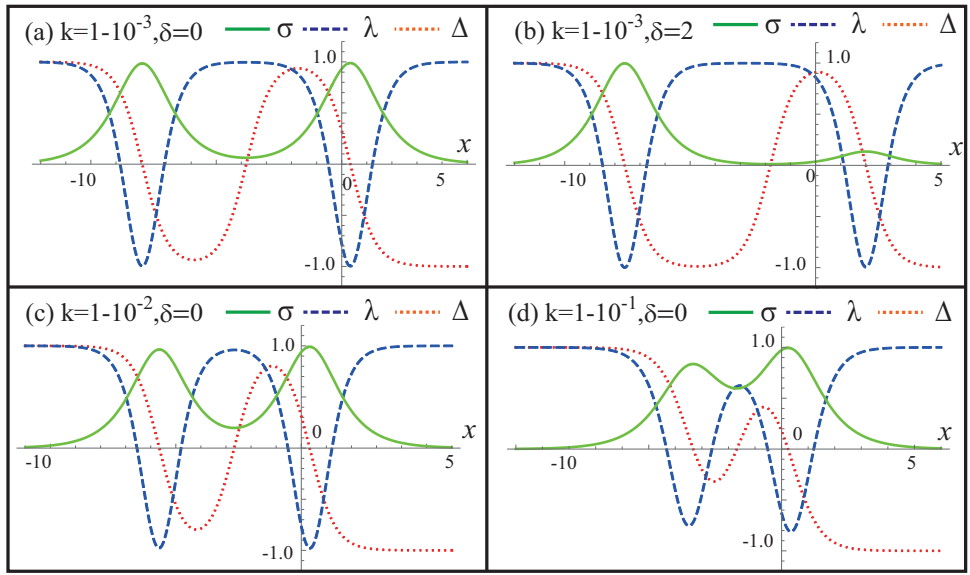}
\caption{The two bright soliton configuration of $\sigma$ (solid line) and $\lambda$ (dashed line). 
The auxiliary field $\Delta$ (dotted line) are also plotted. 
Here we set, $m=1$. 
In Fig.\ (a) and (b) we plot the case of $\delta=0$ and $\delta=2$, respectively, with $k=1-10^{-3}$. 
In Fig.\ (c) and (d) we plot the case of $k=1-10^{-2}$ and $k=1-10^{-1}$, respectively, with $\delta=0$. 
In the figure, we normalize $\sigma$ such that the hight of the highest peak is $1$. 
}
\label{FigThreeKink}
\end{center}
\end{figure}

In the GN model, the integrable structure enables us to 
systematically 
construct all possible 
exact self-consistent solutions \cite{Correa:2009xa,Takahashi:2012aw}. 
The above solutions belong 
to the lowest order 
($n=1$) of the AKNS hierarchy 
(denoted by AKNS$_n$ for $n=1,2,\cdots$) 
for 
the nonlinear Schr\"odinger equation \cite{Correa:2009xa,Takahashi:2012aw} [See Appendix B]. 
The configuration of a kink-anti-kink (polaron)
in the GN model \cite{Campbell:1981dc} 
(in AKNS$_2$) 
does not yield a nontrivial solution in the ${\mathbb C}P^{N-1}$ model, 
while the three kink solution (in AKNS$_3$) 
\cite{Okuno:1983,Feinberg:1996gz,Feinberg:2002nq} 
\begin{align}
\Delta=k \tanh [kx-k\delta+R]
- \frac{ \omega_b e^{R}[\sinh (m_+ x-k\delta+2R)+\sinh(m_-x+k\delta)]}
{\cosh (m_+x-k\delta+2R)+ e^{2R}\cosh(m_-x+k\delta)}, \quad \nonumber
\end{align}
does. 
Here $\omega_b=\sqrt{m^2-k^2}$, $R=(1/2)\ln (m_+/m_-)$, and $m_{\pm}=m\pm k$. 
In Fig.\ \ref{FigThreeKink}, we plot the configurations of $\sigma$, $\lambda$, and $\Delta$ 
for various parameter choices.
The symmetric case $\delta=0$ (a) looks like a double copy of a single soliton in Fig.\ \ref{FigTanh}. 
For larger $\delta$
the middle kink is closer to the right anti-kink than the left anti-kink 
in $\Delta$
as (b), 
and then the amplitude of the Higgs field $\sigma$ localized in the right soliton of $\lambda$ decreases with increasing $\delta$. 
On the other hand,
the parameter $k$ controls the soliton-soliton distance [(a), (c), and (d)]. 
The two solitons merge for larger $k$ and eventually becomes  
one soliton in $k\to 1$.  
This is possible because the three kink solution 
belongs to the same topological sector with the single kink solution 
in the GN model. 
In general, AKNS$_{2k+1}$ ($k=1,2,\cdots$)
yields solutions of $k$ solitons with arbitrary positions  
exhibiting the similar behaviors. 

\section{Summary}
We have found the map from the GN model to the  ${\mathbb C}P^{N-1}$ model, which enables us to construct, for the first time, 
the exact self-consistent stable inhomogeneous solutions 
of the  ${\mathbb C}P^{N-1}$ model; 
a single soliton, a soliton lattice 
and multiple solitons with arbitrary separations. 
The Higgs (broken) phase appears inside the soliton cores where  
the Higgs field $\sigma$ has bright solitonic profiles 
and the ${\mathbb C}P^{N-1}$ moduli are confined.

It is an open question whether there is a map to 
the chiral GN model with continuous chiral symmetry, 
which allows a variety of complex solutions. 
In the (chiral) GN model, the inhomogeneous phase is stabilized 
at the low temperature and high density \cite{Basar:2009fg}, 
or in the presence of a chiral chemical potential, 
equivalent to the constant Zeeman magnetic field on the superconductivity 
\cite{Machida:1984zz}. 
Such analogies in the  ${\mathbb C}P^{N-1}$ model may imply a possibility of a crystaline phase. 
While our periodic soliton lattice can be put on a ring, 
an extension to an interval \cite{Bolognesi:2016zjp,Milekhin:2012ca} 
is also possible to calculate a Casimir force \cite{prep},  
 since the exact solutions in the GN model on an interval 
 have been found recently \cite{Flachi:2017cdo}. 
Another relation between the the ${\mathbb C}P^{N-1}$ model and the GN model in $2+1$ dimensions 
has recently been found in Ref.\ \cite{Filothodoros:2016txa} 
in which the large-$N$ free energy densities for the both theories are found to be remarkably similar. 
Though it would be important to see whether the similar structure also appears in the $1+1$ dimensions, 
we leave it as a future problem. 
The connection between our formalism and the bosonization scheme in $1+1$ dimensions should be also important. 
The former gives the coincidence of the self-consistent equations in ${\mathbb C}P^{N-1}$ model and the GN model, 
whereas the latter yields the sine-Gordon model as the bosonized model of the GN model \cite{Witten:1978qu}. We also leave it as a future problem.
Physical consequences of our solitons on a non-Abelian vortex 
in supersymmetric gauge theories 
\cite{Tong:2005un,Eto:2006pg,Shifman:2007ce,Tong:2008qd} 
or dense QCD \cite{Eto:2013hoa} will be an important problem to be explored.

\section*{Acknowledgement}
The support of the Ministry of Education,
Culture, Sports, Science (MEXT)-Supported Program for the Strategic Research Foundation at Private Universities `Topological Science' (Grant No.\ S1511006) is gratefully acknowledged. 
The work of M.~N.~is 
supported in part by the Japan Society for the Promotion of Science
(JSPS) Grant-in-Aid for Scientific Research (KAKENHI Grant
No.~16H03984) 
and by a Grant-in-Aid for Scientific Research on Innovative Areas ``Topological Materials
Science'' (KAKENHI Grant No.~15H05855) from the MEXT of Japan. 

\appendix

\section{Alternative mapping}
In this Appendix, we show an alternative map from the Gross-Neveu model to the  ${\mathbb C}P^{N-1}$ model. 
In our formalism, $f_n$'s are chosen as upper components of BdG equation $(u_n=f_n,\ v_n=g_n)$ in 
\begin{align}
\left(
\begin{array}{cc}
0 & \partial_x +\Delta \\
-\partial_x+\Delta & 0
\end{array}
\right)
\left(
\begin{array}{c}
u_n \\
v_n
\end{array}
\right)
=
\omega_n
\left(
\begin{array}{c}
u_n \\
v_n
\end{array}
\right),
\end{align}
with 
\begin{equation}
\lambda =\Delta^2+\partial_x \Delta,\ \sigma\propto \exp\left(\int^x dy\Delta\right).
\end{equation} 
For the same $\Delta$, one can also define 
\begin{equation}
\tilde \lambda=\Delta^2-\partial_x\Delta,\ \tilde \sigma\propto\exp\left(-\int^x dy\Delta\right).
\end{equation} 
These functions satisfy 
\begin{align}
&\partial_x^2\tilde \sigma-\tilde\lambda \sigma=0,  \\
&(-\partial_x^2+\tilde \lambda)v_n=\omega_n^2 v_n, \\
&(N/2r)\sum_n u_n v_n=\Delta. 
\end{align}
This implies that the lower component can also be mapped to the ${\mathbb C}P^{N-1}$ model ($v_n=f_n$, $u_n=g_n$) with the Higgs field $\tilde \sigma$ and the mass gap function $\tilde \lambda$. 
Thus the single $\Delta$ corresponds to two solutions in the ${\mathbb C}P^{N-1}$ model (for $\Delta=m$, those are identical). 

For instance, in the case of the kink solution, we obtain 
\begin{align}
&\Delta=m\tanh mx, \\
&\lambda=m^2,\ \sigma=0,\\ 
&\tilde \lambda=m^2(1-2\mathrm{sech}^2 mx),\ \tilde \sigma=A\mathrm{sech} mx,
\end{align}
whereas in the case of the anti-kink solution, we obtain 
\begin{align}
&\Delta=-m\tanh mx, \\
&\lambda=m^2(1-2\mathrm{sech}^2 mx),\ \sigma=A\mathrm{sech} mx,\\ 
&\tilde \lambda=m^2,\ \tilde \sigma=0.
\end{align}
Thus both the solutions correspond to the same solution in the ${\mathbb C}P^{N-1}$ model.

\section{Chiral Gross-Neveu model, Bogoliubov-de Gennes equation, and AKNS hierarchy}
In this Appendix, we briefly summarize the self-consistent treatment of Gross-Neveu model studied in Refs.\ \cite{Correa:2009xa,Takahashi:2012aw}. 
The Lagrangian of the chiral Gross-Neveu model with $N$ flavor is given by 
\begin{equation}
L=\bar\psi i /\hspace{-0.5em}\partial \psi +\frac{g^2}{2}\left[(\bar\psi\psi)^2+(\bar\psi i\gamma^5\psi)^2\right], 
\end{equation}
where $g>0$.
By introducing the auxiliary fields $\Delta_1=-g^2\langle \bar\psi \psi \rangle $ and $\Delta_2=-g^2\langle \bar\psi i\gamma_5 \psi \rangle$, 
and by taking the large $N$ approximation (or mean field approximation) one can obtain the following effective Lagrangian  
\begin{equation}
L_{eff}= \bar\psi i/\hspace{-0.5em}\partial \psi +\left(\Delta_1 \bar\psi\psi+\Delta_2 \bar\psi i\gamma^5\psi\right)
-\frac{1}{2g^2}\left(\Delta_1^2+\Delta_2^2\right). 
\end{equation}
Thus we obtain the following total energy 
\begin{equation}
E_{tot}=\int dx \psi^\dagger H\psi +\frac{1}{2g^2}\int dx (\Delta_1^2+\Delta_2^2), 
\end{equation}
with the Bogoliubov-de Gennes (BdG) Hamiltonian
\begin{align}
H=&-i\gamma^0\gamma^1\frac{d}{dx}-\gamma^0\left(\Delta_1
+ i\gamma^5\Delta_2 \right). 
\end{align}
The consistency condition of the auxiliary field $\Delta_1$ and $\Delta_2$ are called the gap equations  
\begin{equation}
\langle\bar\psi\psi\rangle=-\frac{1}{g^2}\Delta_1,\ \langle\bar\psi i\gamma^5\psi\rangle=-\frac{1}{g^2}\Delta_2, 
\end{equation}
which must be solved in a consistent manner with the BdG equation $H\psi=E\psi$. 
Here the left hand sides of the gap equations can be, respectively, rewritten as 
$N \langle \bar \psi_1 \psi_1\rangle$ and $N\langle \bar\psi_1 i\gamma^5\psi_1\rangle$, 
since the $N$ flavors gives the same contributions, e.g., 
$\langle \bar \psi_1 \psi_1\rangle=\langle \bar \psi_2 \psi_2\rangle=\cdots =\langle \bar \psi_N \psi_N\rangle$. 
Thus we can rewrite the gap equations as 
\begin{equation}
\Delta_1=-g^2N\langle\bar\psi_1\psi_1\rangle,\ 
\Delta_2=-g^2N\langle\bar\psi_1 i\gamma^5\psi_1\rangle. 
\end{equation}
In the following, we use the chiral representation $\gamma_0=\sigma_1$, $\gamma_1=-i\sigma_2$, and $\gamma_5=\sigma_3$. 

For the BdG Hamiltonian, the Gor'kov resolvent $R(x;E)=1/\langle x|(H-E)|x \rangle$ satisfies the Dikii-Eilenberger equation 
\begin{align}
&\partial_x R(x;E)\sigma_3=
\left[
Q(E,\Delta), R(x;E)\sigma_3
\right], \\
&
Q(E,\Delta)=\left(
\begin{array}{cc}
iE&-i\Delta\\
i\Delta^\ast& -iE 
\end{array}
\right), 
\end{align}
where $\Delta=\Delta_1-i\Delta_2$. 
We note that the BdG equation can be written as $\partial_x \psi =Q \psi$. 
The Gor'kov resolvent must satisfies the conditions $\det R=-\frac{1}{4}$, $\mathrm {Tr} R\sigma_3=0$, and $R^\dagger =R$.  

The Dikii-Eilenberger equation and the BdG equation can be rewritten as 
\begin{align}
&\partial_t Q-\partial_x R\sigma_3+
\left[
Q, R\sigma_3
\right]=0,\ 
\partial_x \psi = Q\psi,\ 
\partial_t \psi = R\sigma_3\psi, 
\end{align}
with the constraint $\partial_t Q=0$. 
The first equation is the integrable condition (zero curvature condition) of this system; $\partial_x\partial_t \psi =\partial_t\partial_x\psi$. 
Since we find the connection between BdG system to 
the AKNS system, by using the machinery of the integrable system, 
one can systematically expand 
the resolvent $R\sigma_3$ which yields AKNS$_n$ as 
\begin{equation}
R\sigma_3=i\sum_{j=1}^{n+2} c_j V^{(j)},\ V^{(n)}=\sum_{j=0}^{n-1} (2E)^{n-1-k}M^{(j)}, 
\end{equation}
where $c_j$'s are positive constants. 
Here $M^{(i)}_{i, j}$ components of the matrices $M^{(i)}$ satisfy $M_{11}^{(i)}=-M_{22}^{(i)}$, $M_{12}^{(i)}=(M_{21}^{(i)})^\ast$, and first few components are given by 
\begin{align}
&M_{11}^{(0)}=-\frac{i}{2}, 
M_{12}^{(0)}=0,
\\ 
&M_{11}^{(1)}=0, 
M_{12}^{(1)}=i\Delta, 
\\ 
&M_{11}^{(2)}=-i|\Delta|^2, 
M_{12}^{(2)}=\partial_x \Delta,
\\ 
&M_{11}^{(3)}=-2i \Im (\Delta^\ast\partial_x  \Delta), 
M_{12}^{(3)}=\partial_x^2\Delta-2|\Delta|^2,\\
&M_{11}^{(4)}=2i\Re (\Delta^\ast \partial_x^2\Delta)-2i |\partial_x\Delta|^2-3|\Delta|^4, \nonumber\\
&M_{12}^{(4)}=-\partial_x^3\Delta+6|\Delta|^2\partial_x\Delta.
\end{align}
The higher components are calculable with a help of the following formula 
\begin{equation}
\frac{i}{2}\left[\sigma_3, M^{(n+1)}\right]=\partial_x M^{(n)}+\left[M^{(1)}, M^{(n)}\right]. 
\end{equation}
We can also obtain the nonlinear Schr\"odinger equations for this system as $\sum_{j=1}^{n+1} c_j M^{(j)}_{12}=0$. 
The AKNS$_0$, AKNS$_1$, AKNS$_2$ for instance, yield 
\begin{align}
&-\frac{i}{2}\partial_x \Delta+c_1\Delta=0,\\ 
&-\frac{1}{4}(\partial_x^2 \Delta-2|\Delta|^2\Delta)-c_1\frac{1}{2}\partial_x\Delta+c_2\Delta=0,\\
&\frac{i}{8}(\partial_x^3 \Delta-8|\Delta|^2\partial_x^2\Delta)
-c_1\frac{1}{4}(\partial_x^2 \Delta-2|\Delta|^2\Delta)-c_2\frac{1}{2}\partial_x\Delta+c_3\Delta=0.  
\end{align}
The fermionic solutions are also calculable as 
\begin{align}
&\psi_1^2=CV_{12}\sqrt{\frac{iV_{11}-\omega}{iV_{11}+\omega}}\exp\left[i\omega \int^x_0dx\left(\frac{U_{12}}{V_{12}}+\frac{U_{21}}{V_{21}}\right)\right],\\ 
&\psi_1^2=-CV_{21}\sqrt{\frac{iV_{11}+\omega}{iV_{11}-\omega}}\exp\left[i\omega \int^x_0dx\left(\frac{U_{12}}{V_{12}}+\frac{U_{21}}{V_{21}}\right)\right],
\end{align}
where $\psi=(\psi_1, \psi_2)^T$ and $C$ is the normalization constant. The square-root of those function must be taken such as $v/u=iV_{21}/(iV_{11}-\omega)$.

\end{document}